# Landscape of IoT Patterns

Hironori Washizaki
*Waseda University / NII / SYSTEM INFORMATION / eXmotion,* Tokyo, Japan
washizaki@waseda.jp

Nobukazu Yoshioka
*National Institute of Informatics*
Tokyo, Japan

Atsuo Hazeyama
Tokyo Gakugei University
Tokyo, Japan

Takehisa Kato
Toshiba Digital Solutions Corporation
Kanagawa, Japan

Haruhiko Kaiya
Kanagawa University
Kanagawa, Japan

Shinpei Ogata
Shinshu University
Nagano, Japan

Takao Okubo
Institute of Information Security
Kanagawa, Japan

Eduardo B. Fernandez
Florida Atlantic University
Boca Raton, USA

*Abstract*—Patterns are encapsulations of problems and solutions under specific contexts. As the industry is realizing many successes (and failures) in IoT systems development and operations, many IoT patterns have been published such as IoT design patterns and IoT architecture patterns. Because these patterns are not well classified, their adoption does not live up to their potential. To understand the reasons, this paper analyzes an extensive set of published IoT architecture and design patterns according to several dimensions and outlines directions for improvements in publishing and adopting IoT patterns.

*Keywords—Patterns, Internet of Things (IoT), Design, Architecture, Survey, Systematic Literature Review (SLR)*

## I. INTRODUCTION

The Internet of Things (IoT) aims to bring connectivity to almost every object (i.e., things found in physical space). Although it extends the connectivity to everyday things, such an increase in connectivity creates many challenges [1]. Patterns are encapsulations of reusable common problems and solutions under specific contexts. As the industry is realizing many successes (and failures) in IoT systems development and operations, many IoT patterns have been published such as IoT design patterns and IoT architecture patterns. However, these patterns are not well classified. Consequently, their adoption does not live up to their potential.

The contributions of this paper are an overview of the current landscape of the IoT architecture and design patterns, identification of shortcomings, and suggestions to improve publishing and adoption of IoT patterns. Here, a complete (to the authors' knowledge) set of IoT patterns that is available in the literature is analyzed. The authors surveyed 33 papers published from 2014–2018. For each question below, the directions for improvement are outlined constructively.

**RQ1.** *What are the publication trends of IoT patterns?* To answer this question, we identified the publication years and venues of the 33 papers surveyed.

**RQ2.** *Are all existing IoT patterns really IoT patterns?* To answer this question, we confirmed whether or not each proposed or used pattern the IoT context is actually a pattern addressing specific problems and solutions in IoT.

**RQ3.** *Do the IoT pattern appropriately cover the issues in IoT development and operations?* To answer this question, we built a classification scheme for IoT patterns and classified each IoT pattern based on the scheme, which includes abstraction level, domain specificity, and quality characteristics.

The rest of the paper is structured as follows. Section II presents the classification scheme. Section III introduces the main sources and analysis process used in this study. Section IV analyzes the literature according to the above-mentioned scheme and process. Finally, Section V presents the conclusions and future work.

## II. IoT PATTERN CLASSIFICATION

To classify IoT patterns, we identified three dimensions: abstraction level, domain specificity, and quality characteristic.

### A. Abstraction Level

In general, the IoT systems development process has several major phases with abstraction levels. From the most to the least abstract level, they are analysis, system and software architecture design, detail design and construction. According to these phases, IoT design patterns can be classified into the following three types in terms of abstraction level.

1) High: Reference architectures are architectural models that specify architectural elements and their connections at a very high abstraction level without much consideration of concrete platforms or detailed design issues. These are often used at early phases such as analysis and architecture design.

2) Middle: Unlike reference architectures, there are recommended concrete architecture design of IoT systems and software to address repeated concrete architectural problems such as ensuring interoperability among heterogeneous devices. These architectures are often documented as architecture patterns that encapsulate contexts, recurrent problems and their corresponding solutions. We regarded the abstraction level of the architecture patterns as between high and low.

3) Low: There are recommended detailed designs to address recurrent detailed design problems such as enabling proper communications among software modules while keeping a high extensibility. Since these patterns target specific modules or limited parts of entire system and software, and not the entire software and systems, we regarded the abstraction level of the design patterns as low. These are often used at detail design and construction phases.

*B. Domain Specificity*

Domain specificity is important to examine the applicability and reusability of each IoT pattern. It is divided into three types: any, general IoT and specific IoT.

1) Any: General systems and software architecture patterns as well as design patterns that can be adopted to design IoT systems and software if their contexts and problems match the patterns' contexts and problems.

2) General IoT: IoT architecture and design patterns, which are applicable to any IoT systems and software.

3) Specific IoT: IoT architecture and design patterns that address specific problem domains (such as the healthcare) and technical domains (such as the brain-computer interaction).

*C. Quality Characteristic*

All systems and software design patterns address some quality characteristics. Basically, IoT design patterns should address interoperability, which is defined as a sub-characteristic of compatibility in ISO/IEC 25010:2010 [2]. We use all quality characteristics except for the functional suitability defined in ISO/IEC 25010, which is a well-accepted quality model systems and software engineering. Additionally, there are other emerging characteristics that are not defined in ISO/IEC 25010 but are common in IoT development and operation. Possible candidates are scalability and privacy.

III. ANALYSIS PROCESS

We use a systematic literature review (SLR) to evaluate relevant publications on IoT patterns. A SLR aims to assess scientific papers to group concepts around a topic.

1) Initial Search: We used Scopus (https://www.scopus.com/) as a well-accepted reliable scientific databases and indexing systems. For consistency, we executed the following query on titles, abstracts, and keywords of papers regardless of time and subject area. We found 63 papers published from 2014–2018.

"IoT" AND ( "design pattern" OR "architecture pattern" )

2) Impurity Removal: Due to the nature of the involved data source, the search results included some elements that are clearly not research papers such as abstracts and international standards. By removing these results, we got 56 papers.

3) Inclusion and Exclusion Criteria: Applying the following criteria reduced the number of papers to 33 [3-35].

- Inclusion: Papers addressing patterns for designing IoT systems and software, and papers written in English
- Exclusion: Papers that focus on IoT but do not explicitly deal with architecture and design patterns, and papers that are duplicates of other studies

4) Data Extraction: The following information was collected from each paper to answer the research questions: Publication title, publication year, publication venue, types of patterns proposed or used, pattern names, domain names in the case of Specific IoT patterns, and quality characteristics addressed.

IV. RESULT AND DISCUSSION

*A. Publication (RQ1)*

Table 1 presents the distribution of publications over time. The most common publication types are conference papers (17), journals (8), workshops (5), symposiums (2), and refereed book chapter (1). The high numbers of conference papers and journal papers suggest IoT architecture and design patterns are maturing. Since 2016, IoT patterns have become an important and eye-catching aspect of research, and interest has been expanding each year.

Table 1. Primary studies by publication type and year

|  | Workshop | Symposium | Conference | Book chapter | Journal | Total |
|---|---|---|---|---|---|---|
| 2014 |  |  | 1 |  |  | 1 |
| 2015 |  |  | 1 |  |  | 1 |
| 2016 |  | 1 | 3 |  | 2 | 6 |
| 2017 | 1 | 1 | 7 |  | 3 | 12 |
| 2018 | 4 |  | 5 | 1 | 3 | 13 |
| Total | 5 | 2 | 17 | 1 | 8 | 33 |

*B. IoT Patterns (RQ2)*

We identified 136 patterns mentioned in 33 papers. Among them, 75 patterns (55%) are classified as "Any" in terms of domain specificity. Major non-IoT patterns that appear in multiple papers include *Publish-Subscribe* [6,7,24,33,34], *Client-Server* [24,33,34], *Peer-to-Peer* [24,33], *REpresentational State Transfer (REST)*[33,34], *Service Oriented Architecture (SOA)*[24,34], *Role Based Access Control (RBAC)*[12,15], *Model-View-Controller (MVC)*[20,28], and *Reflection*[8,27]. The other 67 patterns appear in one paper only. Surprisingly, 14 papers used such non-IoT patterns only. According to these results, we confirmed that IoT systems and software are often designed via conventional architecture and design patterns that are not specific to IoT design.

There are 61 IoT patterns (i.e., 45%) in 19 papers that address specific problems and solutions in IoT. The details are discussed in the subsequent section.

*C. IoT Pattern Classification (RQ3)*

Table 2 presents the distribution of IoT and non-IoT patterns by abstraction level and domain specificity. Table 3 presents the list of 61 IoT architecture and design patterns.

Surprisingly, only one pattern *Operator-Controller-Module (OCM)* is mentioned in multiple papers [4,32]. The remaining appear once in different papers. This indicates that IoT patterns are not shared or recognized by different research groups. This may be due to its short history. Potential pattern authors are encouraged to carefully check existing IoT patterns before publishing their own "new" patterns.

In terms of abstraction level, about half of 61 IoT patterns are IoT architecture patterns (i.e., 48%), 27 cover IoT design patterns (44%), and 5 reference architectures (8%). We confirmed that IoT architecture patterns and IoT design patterns are almost equally proposed or used.

In terms of domain specificity, 41 patterns (i.e., 67%) are general IoT, while the remaining 20 patterns (33%) are specific to a problem or technical domain shown in Table 3.

Reviewing the combinations of abstraction level and domain specificity, most of IoT design patterns are applicable to any domain but many IoT architecture patterns exist for specific domains. This implies that the unique nature of IoT adoption in specific domains often appears at the architecture level. Design details seem to be commonly addressed by general IoT design patterns or even non-IoT design patterns. In the future, the number of specific IoT design patterns may increase as more domains adopt IoT.

In terms of quality characteristics, many IoT patterns address performance efficiency, compatibility (including interoperability as a sub-characteristic), usability, reliability, and maintainability. This finding is quite natural since major concerns in IoT adoption revolve around these characteristics. Consequently, other quality characteristics remain to be researched. A few number of IoT patterns address security, portability (including adaptability as a sub-characteristic) and scalability. Privacy is rarely addressed by IoT patterns.

Table 2. Patterns by abstraction level and domain specificity

| Abstraction level | Domain specificity | | | Total |
|---|---|---|---|---|
| | Any | General IoT | Specific IoT | |
| High | 20 | 3 | 2 | 25 |
| Middle | 14 | 14 | 15 | 43 |
| Low | 41 | 24 | 3 | 68 |
| Total | 75 | 41 | 20 | 136 |

*D. Limitations*

The classification of the patterns was conducted by all authors except for the last author of this paper and reviewed by the first author. It is possible that our classification results may not be completely correct. To mitigate this threat to validity, we will open the classification results to the public and call for comments at our Website.

We used Scopus as the initial document base of the SLR. Although it is adopted in other SLRs, relevant papers may be missed. To mitigate this threat, we plan to use other databases, extend our SLR, and elicit public review of the revised results.

V. CONCLUSION AND FUTURE WORK

To overview the current landscape of IoT architecture and design patterns, we surveyed about 136 patterns mentioned in 33 papers published between 2014–2018 according to several dimensions. Most of IoT design patterns are applicable to any domain but many IoT architecture patterns exist for specific domains. In the future, the number of specific IoT design patterns may increase as more domains adopt IoT. In terms of quality characteristics, many IoT patterns address performance efficiency, compatibility, usability, reliability, and maintainability. Consequently, other quality characteristics remain to be researched. Our future works include further analysis on IoT patterns using additional dimensions such as relationships among patterns and writing quality of patterns.

Table 3. Detailed list of IoT patterns (Pe: Performance efficiency, C: Compatibility, U: Usability, R: Reliability, Se: Security, M: Maintainability, Po: Portability, Sc: Scalability, Pr: Privacy)

| Abstraction | Specificity | Pattern | Domain | Pe | C | U | R | Se | M | Po | Sc | Pr | Paper |
|---|---|---|---|---|---|---|---|---|---|---|---|---|---|
| High | General | Layered architecture for IoT applications | | X | | | X | X | | | X | X | [29] |
| High | General | Lambda-style architecture | | X | | | X | | X | X | X | | [31] |
| High | General | Kappa-style architecture | | X | | | X | | X | X | X | | [31] |
| High | Specific | Security Architecture (for Smart Water Management) | Smart Water Management | | | | | X | X | | X | | [7] |
| High | Specific | Machine intelligence layer for industrial IoT | Machine Intelligence | | X | | | | | | X | | [35] |
| Middle | General | Alignment-based Translation | | | X | | | | X | | | | [14] |
| Middle | General | AS2AS Discovery of IoT Services | | | X | | | | X | | | | [14] |
| Middle | General | AS2AS Flow-based Service Composition | | | X | | | | X | | | | [14] |
| Middle | General | AS2AS Service Orchestration | | | X | | | | X | | | | [14] |
| Middle | General | D2D REST Request/Response | | | X | | | | X | | | | [14] |
| Middle | General | IoT artifact's Middleware Message Broker | | | | | | | X | X | | | [14] |
| Middle | General | IoT Artifact's Middleware Message Translator | | | X | | | | X | | | | [14] |
| Middle | General | IoT Artifact's Middleware Self-contained Message | | | | | | | X | | | | [14] |
| Middle | General | IoT Artifact's Middleware Simple Component | | | | | | X | X | | | | [14] |
| Middle | General | IoT Gateway Event Subscription | | | X | | | | X | | | | [14] |
| Middle | General | Orchestration of SDN Network Elements | | | X | | | | X | | | | [14] |
| Middle | General | IoT SSL CROSS-Layer Secure Access | | | X | | | X | X | | | | [14] |
| Middle | General | Translation with Central Ontology | | | X | | | | X | | X | | [14] |
| Middle | General | Entity-Component-Attribute | | | X | | | | X | X | | | [17] |
| Middle | Specific | Blockchain-based Architecture | Trusted Orchestration Management | X | | | X | X | X | | X | | [9] |
| Middle | Specific | Closed-Loop: Classical Closed-Loop Control | Industrial IoT | X | | | X | | | | | | [16] |
| Middle | Specific | Cloud-in-the-Loop: Closed-Loop Control via the Cloud | Industrial IoT | X | | | X | | | | | | [16] |
| Middle | Specific | Cloud-on-the-Loop: Cloud-configured Control | Industrial IoT | X | | | X | | | | | | [16] |
| Middle | Specific | Device-to-Device (D2D): Local Coordination | Industrial IoT | X | | | X | | | | | | [16] |
| Middle | Specific | Open-Loop: Classical Open-Loop Control | Industrial IoT | X | | | X | | | | | | [16] |
| Middle | Specific | Publisher: Sensor Data Publication | Industrial IoT | | X | | | | X | | | | [16] |
| Middle | Specific | Design pattern for computation offloading | Computation Offloading | X | | | | | X | X | | | [18] |
| Middle | Specific | Landline Interception | Emergency Information Delivery | | | X | | | | | | | [21] |
| Middle | Specific | SIM Equipped Device | Emergency Information Delivery | | | X | | | | | | | [21] |
| Middle | Specific | SMS to Display over Bluetooth/Wi-Fi | Emergency Information Delivery | | | X | | | | | | | [21] |
| Middle | Specific | SMS to Mobile Application | Emergency Information Delivery | | | X | | | | | | | [21] |
| Middle | Specific | Web System (for Emergency Information Delivery) | Emergency Information Delivery | | | X | | | | | | | [21] |
| Middle | Specific | Actor | Blockchain | | X | | | | X | | | | [25] |
| Middle | Specific | Operator-Controller-Module (OCM) | Organic Rankine Cycle Turbine | X | | | | | X | | | | [4, 32] |
| Low | General | Edge Code Deployment | | | | | | | | X | X | | [19] |
| Low | General | Edge Diameter of Things (DOT) | | | | | | | | X | | | [19] |
| Low | General | Edge Orchestration | | | | | X | | | X | X | | [19] |
| Low | General | Edge Provisioning | | | | | X | | | X | | | [19] |
| Low | General | Frame Buffer | | X | | | X | | | | | | [22] |
| Low | General | Slot Buffer | | X | | | X | | | | | | [22] |
| Low | General | Application launch | | | | X | | | | | | | [23] |
| Low | General | Get details of a device | | | | X | | | | | | | [23] |
| Low | General | Get Information for one category | | | | X | | | | | | | [23] |
| Low | General | Get information from the device | | | | X | | | | | | | [23] |
| Low | General | Get state of the device | | | | X | | | | | | | [23] |
| Low | General | More devices more operations | | | | X | | | | | | | [23] |
| Low | General | More devices one operation | | | | X | | | | | | | [23] |
| Low | General | Nearby devices | | | | X | | | | | | | [23] |
| Low | General | One category more operations | | | | X | | | | | | | [23] |
| Low | General | One device more operations | | | | X | | | | | | | [23] |
| Low | General | One device one operation | | | | X | | | | | | | [23] |
| Low | General | One device one program | | | | X | | | | | | | [23] |
| Low | General | Pull information | | | | X | | | | | | | [23] |
| Low | General | Push information | | | | X | | | | | | | [23] |
| Low | General | Search device | | | | X | | | | | | | [23] |
| Low | General | Action interaction | | | | | | | X | | | | [30] |
| Low | General | Event interaction | | | | | | | X | | | | [30] |
| Low | General | Property interaction | | | | | | | X | | | | [30] |
| Low | Specific | Ontology Design Pattern for IoT Device Tagging Systems | Building Automation | | X | | | | X | | | | [10] |
| Low | Specific | Actuation-Actuator-Effect (AAE) ontology design | Brain-Computer Interaction | | X | | | | X | | | | [13] |
| Low | Specific | Stimulus-Sensor-Observation (SSO) ontology design | Brain-Computer Interaction | | X | | | | X | | | | [13] |
| | | Number of patterns that address the corresponding quality characteristic | | 13 | 17 | 20 | 14 | 6 | 31 | 6 | 8 | 1 | |